%% file: main.tex
\documentclass[twocolumn]{aastex61}
\pdfoutput=1 
\usepackage{amsmath,amstext}
\usepackage[T1]{fontenc}
\usepackage{apjfonts} 
\usepackage[figure,figure*]{hypcap}
\usepackage{subfigure}

\shorttitle{Asteroid Rotation Periods in TESS}
\shortauthors{McNeill et al.}

\begin{document}

\title{Asteroid Photometry from the Transiting Exoplanet Survey Satellite: A Pilot Study}

\author{A. McNeill}
\affiliation{Department of Physics and Astronomy, Northern Arizona University, Flagstaff, AZ 86011, USA}
\author{M. Mommert}
\affiliation{Lowell Observatory, 1400 W.\ Mars Hill Rd., Flagstaff, AZ, 86001, USA}
\author{D.E. Trilling}
\affiliation{Department of Physics and Astronomy, Northern Arizona University, Flagstaff, AZ 86011, USA}
\affiliation{South African Astronomical Observatory, 
PO Box 9,
7935 Observatory
South Africa}
\author{J. Llama}
\affiliation{Lowell Observatory, 1400 W.\ Mars Hill Rd., Flagstaff, AZ, 86001, USA}
\author{B. Skiff}
\affiliation{Lowell Observatory, 1400 W.\ Mars Hill Rd., Flagstaff, AZ, 86001, USA}

\begin{abstract}

The {\it Transiting Exoplanet Survey Satellite} (TESS) searches for planets transiting bright and nearby stars using high-cadence, large-scale photometric observations. Full Frame Images provided by the TESS mission include large number of serendipitously observed main-belt asteroids.
Due to the 
cadence of the published Full Frame Images
we are sensitive to periods as long as of order tens of days, a region of phase space that is generally not accessible through traditional observing. This work represents a much less biased measurement of the period distribution in this period range.
We have derived rotation periods for 300~main-belt asteroids and have partial lightcurves for a further 7277 asteroids,
including 43 with periods $P > 100$ h;
this large number of slow rotators is predicted by theory.
Of these slow rotators we find none requiring significant internal strength to resist rotational reshaping.
We find our derived rotation periods to be in excellent agreement with
results in the Lightcurve Database for the 55~targets that overlap.
Over the nominal two-year lifetime of the mission, we expect the detection of around 85,000 unique asteroids with rotation period solutions for around 6000 asteroids.
We project that the systematic analysis of the entire TESS data set will increase the number of known slow-rotating asteroids (period > 100~h) by a factor of 10.
Comparing our new period determinations with previous measurements in the literature,
we find that the rotation period of
asteroid 
(2320) Blarney  has
decreased by at least 20\% over the past
decade, potentially due to surface activity or subcatastrophic collisions.
\end{abstract}


\section{Introduction}

\begin{figure}
  \begin{center}
\includegraphics[width=0.5\textwidth]{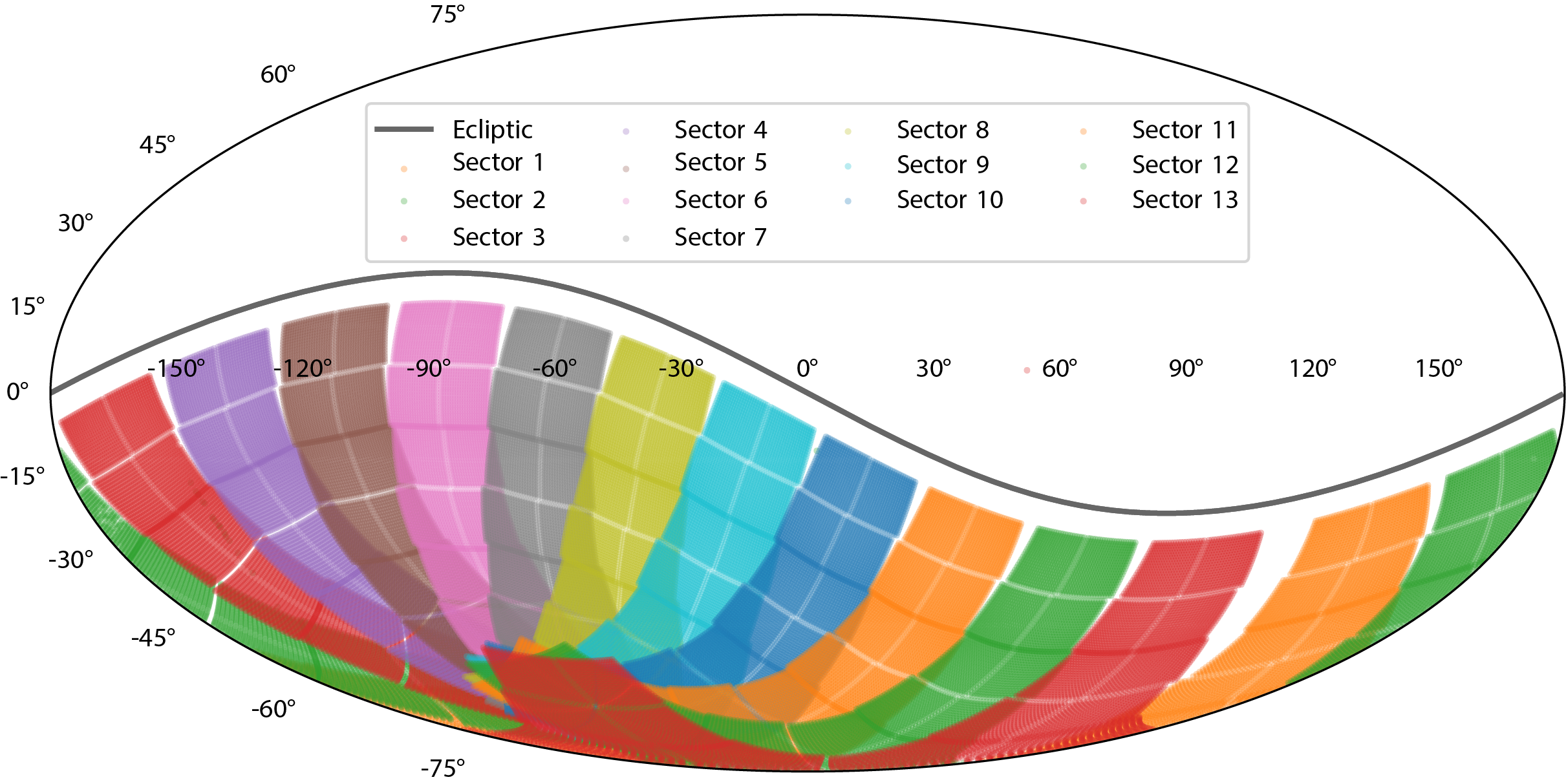}   
\caption{A diagram in celestial coordinates showing the pointings for the first year of the TESS mission. In the second year of operations TESS will observe the Northern Hemisphere.}
\label{tess}
  \end{center}
 \end{figure}

TESS, the {\it Transiting Exoplanet Survey Satellite} (\citealt{ricker2014}), was launched in April 2018 with the goal of finding Earth-like exoplanets that transit nearby stars. TESS will observe the entire sky in its nominal two year mission.
The details of TESS' observing cadence and geometry provide a unique opportunity to measure the lightcurves of thousands of asteroids continuously for 
up to 
twenty-seven days as
those objects pass through the TESS field of view (Figure~\ref{tess}).
This large-area survey allows us unbiased sensitivity to rotation periods from hours (limited by the 30~minute reporting cadence) to many days (limited by the 27~day staring time). This is the first large-scale data set to allow a probe of long asteroid rotation periods for many thousands of asteroids. This in turn allows measurements of the strength properties of asteroids, which constrain details of their formation.

We have carried out a pilot study based on a small sample of the initial TESS public data release. Here we report rotation periods
for 300~main-belt asteroids and
detections or partial lightcurves for a further 7277 asteroids.
In Section~2 we provide background on the importance of asteroid lightcurves, and particularly long-period lightcurves, for understanding the formation of small bodies in our Solar System. In Section~3 we give an overview of the TESS observing cadence.
Section~4 presents our technique of extracting asteroid photometry from TESS data.
In Section~5 we present our resulting asteroid lightcurves and in Section~6 we detail our future planned work through the lifetime of the TESS mission.


\section{Asteroid Rotations}

Approximately $95\%$ of all asteroids are found in the main asteroid belt
between Mars and Jupiter. These minor planets represent remnant material from the initial formation of the solar system. 
Understanding the physical properties and evolution of asteroids helps inform models aiding our understanding of planetary system formation. Since their initial formation the rotational behaviour of asteroids can have undergone considerable evolution. The main factors affecting the evolution of rotation properties are collisions between asteroids, tidal interactions with large bodies, internal damping, and the Yarkovsky-O'Keefe-Radzievskii-Paddack (YORP) Effect. A clear picture of the rotational properties of asteroids will lead to increased understanding of these processes and hence collisional and dynamical evolution of the main asteroid belt. 
 
 Most objects for which rotation periods have been derived have relatively short periods ($< 1$ day). This is an observational bias as obtaining accurate rotation periods for slow rotators is highly resource intensive. For this reason most long light curve periods determined for asteroids tend to have been observed by sky surveys and their periods determined from sparse data. 
 
Just as an object which rotates quickly may need some internal strength to maintain its shape,
a highly elongated
slowly rotating asteroid also requires strength to resist 
either mass loss or becoming more spherical. Therefore, elongated slowly rotating asteroids require finite strength. The difficulty in obtaining accurate rotation periods for these objects additionally presents difficulties in obtaining an unbiased period distribution which is needed for shape and spin distribution models.

\begin{figure}
  \begin{center}
\includegraphics[width=0.5\textwidth]{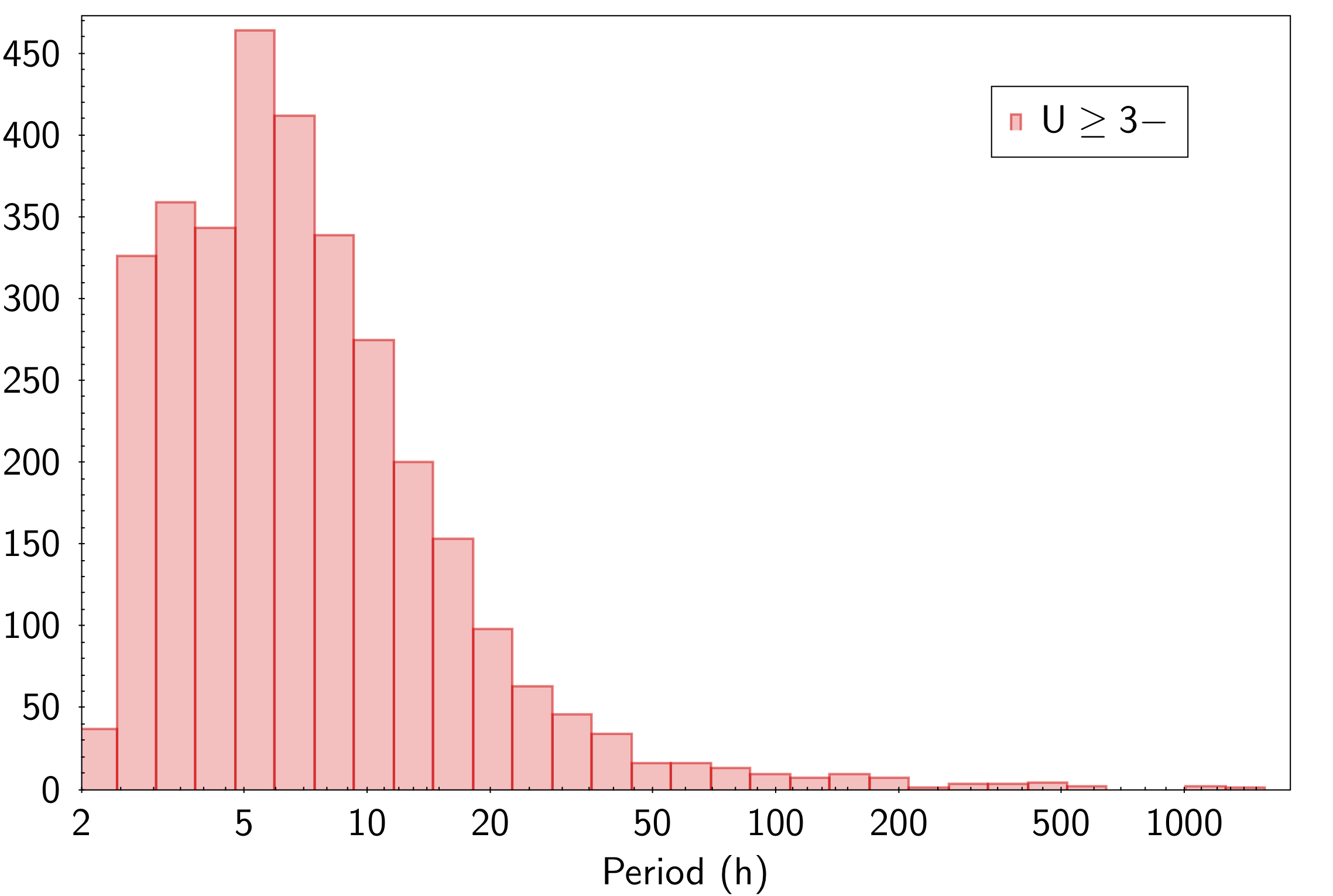}   
\caption{A histogram showing the distribution of rotation periods for asteroids stored in the LCDB with assigned quality value $U \geq 3-$.
There are currently only 58 asteroids with well-constrained
periods longer than 100~hours because of the resources needed to detect and confirm these slow-rotators with conventional ground-based
nighttime observing.}
\label{lcdb_histogram}
  \end{center}
 \end{figure}

Figure~\ref{lcdb_histogram} shows objects for which rotation periods are listed in the Light Curve Database (\citealt{warner2009}). The quality factor, $U$, is assigned to a light curve based on the confidence with which the rotation period was determined. $U$ values range from $U=1$ where the period is likely to be incorrect to $U=3$ where the rotation period is unambiguous. The numerical quality code in some cases is followed by a positive or negative, showing that the reliability is judged slightly better or worse than implied by an unsigned number alone.  Figure~\ref{lcdb_histogram} shows the distribution of rotation periods for objects $U \geq 3-$. This shows the dearth of measured rotation periods for slow-rotating asteroids ($P > 100$ h). The resource intensive nature of observing slow rotators means that those objects are generally observed at a sparse cadence. This leads to generally uncertain period solutions, which
explains the small number of good solutions for periods greater than 50~hours.

The spin rate distribution of main belt asteroids is not consistent across all sizes. The distribution of rotation rates for large diameter objects (D > 40 km) takes the form of a Maxwellian distribution (\citealt{pravec2000}). The spin rate distribution of objects smaller in size is non-Maxwellian. Compared to large diameter asteroids there is an excess of both fast and slow rotators in this size range, possibly as a result of the YORP effect (\citealt{bottke2006}). Through this work we hope to obtain a more complete sample of these slow rotators in the main belt population. 



\section{The {\it Transiting Exoplanet Survey Satellite}}

The primary science goal of 
TESS is the search for exoplanets transiting bright and nearby stars (\citealt{ricker2014}). The satellite is situated in a highly elliptical 13.7-day orbit around the Earth. TESS observes with four wide-field optical cameras (each consisting of four 2k$\times$2k CCDs) with a total field of view for each sector of 24$\times$90 degrees, with 13 sectors (pointings) in the southern ecliptic hemisphere scheduled for Year 1 and a further 13 in the northern hemisphere in Year 2.
Each pointing will be observed for an interval ranging from one month to one year, with the longest observing epochs near the ecliptic poles, which are continuously observed through the different pointings,  
Individual frames are taken every two minutes and stacked into Full Frame Images covering 30 minutes. The 
point-source 
magnitude range of the Full Frame Images corresponds approximately to $8 < V < 17$,
with 21~arcsecond pixels.

Whilst the primary mission of the satellite is the study of exoplanets, over the course of the mission lifetime tens of thousands of main belt asteroids will enter the field of view, primarily in Camera 1, which points closest to the ecliptic. This provides an opportunity for an untargeted survey of main belt asteroids, whose sky motions and 
magnitudes are well suited to TESS observations.
These data will be particularly valuable to determine rotation periods and lower-limit elongations for a large population of asteroids. The study of sparse photometry for an untargeted survey has been carried out in the past using ground-based telescopes (\citealt{mcneill2016}, \citealt{cibulkova2018}). The key difference between these past studies and data from TESS is the space-based nature of the satellite, allowing for much longer baseline observations of asteroids without interruption by daylight. This, along with the long intervals the survey points at a given sector, makes this a unique opportunity to obtain observations of slow-rotating asteroids --- bodies with periods longer than 100~hours, which cannot be easily observed with classical observing and ground-based telescopes.

\section{Extracting Photometry from TESS}

We downloaded the publicly available calibrated Full Frame Images of TESS sectors 1 and 2 using the bulk download services provided as the Barbara A. Mikulski Archive for Space Telescopes (MAST)\footnote{MAST bulk download scripts: \url{http://archive.stsci.edu/tess/bulk_downloads/bulk_downloads_ffi-tp-lc-dv.html}}. Full Frame Images are stored based on their sector and camera identifiers; each downloaded FITS data cube is split into 4 separate flat FITS files according to their CCD identifiers. From each of these CCD frames the background has been subtracted. The background has been measured as the median derived in a 100$\times$100 pixel cell and its eight neighboring cells utilizing a 3$\sigma$ sigma-clipping rejection ({\tt astropy.photutils}).

From all background-subtracted CCD frames for the same sector-camera-CCD combination, 50 frames are selected that are evenly distributed in time to generate a median background template for this sector-camera-CCD combination. The even spacing across time is necessary to capture systematic changes in the telescope pointing or temporally changing artifacts due to nearby bright sources. We subtract the template image from each frame with the same sector-camera-CCD combination in order to remove fixed sources from the frame. Due to the aforementioned systematic changes, image artifacts are introduced, mainly around bright sources. In order to minimize the effect of these artifacts on the resulting photometry, we create an image mask that covers these artifacts in each frame. The image mask is produced by convolving a 5$\times$5 pixel mask with those pixels that have 
values less than the local median background minus 3 times the local background rms around the pixel location. The size of the pixel mask is chosen to conservatively mask the remainder flux from insufficiently subtracted background sources.

\subsection{Aperture Photometry}

Target magnitudes are measured with forced aperture photometry using {\tt astropy.photutils}. The target flux is measured in a circular aperture with a radius of 2 pixels. This size has been chosen to include a significant fraction of the target flux (a 2$\times$2 pixel aperture centered on a point source includes $90\%$ of the total target flux\footnote{\url{https://heasarc.gsfc.nasa.gov/docs/tess/the-tess-space-telescope.html}}). The background is measured as the median pixel value within an annulus with an inner radius of 4 pixels and an outer radius of 8 pixels. Masked pixels are ignored in the photometric measurement. Uncertainties are estimated as the geometric sum of the target's Poisson noise in units of photons and the background noise across the target aperture and the annulus that was used to estimate the sky background (this method is identical to the one used by the IRAF {\tt noao.digiphot.apphot.phot} method\footnote{\url{http://stsdas.stsci.edu/cgi-bin/gethelp.cgi?phot}}). 

The resulting photometry is written to a file on a per-object basis. In addition to positions and photometric properties, a flag indicates whether the aperture is affected by a pixel mask, enabling the quick identification of unreliable photometric measurements.

\subsection{Target Identification}

Asteroids in the CCD frames were identified in a two-step process. In the first step, candidate asteroids that could be present in any CCD of the sector-camera combination are identified. For this purpose, ephemerides of all known asteroids are calculated for the midtime epoch of all available observations for this sector. Any asteroid within a 25\degr radius of the sector-camera center position and with a visual magnitude brighter than 20 is considered a candidate. Orbit calculations are performed using OpenOrb (\citealt{granvik2009}) via the Python extension {\tt pyoorb}\footnote{\url{https://github.com/oorb/oorb/tree/master/python}} and {\tt sbpy.data.Orbit} \citep{mommert2019}. Orbital elements used in this calculation are obtained from the Minor Planet Center orbit file {\tt MPCORB.DAT}\footnote{\url{https://minorplanetcenter.net/iau/MPCORB.html}} file as of 8 Jan.\ 2019. Ephemerides are calculated relative to the geocentric location, which introduces uncertainties in the positions of near-Earth asteroids, but does not significantly affect main belt asteroids. Additional uncertainty is introduced by the offset of the orbital element epoch from the actual epoch of the observations. In order to compensate for these uncertainties, a search cone (radius of 25\degr) much larger than the actual field of view (radius of $\sqrt{2}\cdot 12$\degr) has been chosen. This approach was chosen over a more accurate ephemerides calculation in order to improve the computational performance of this search. 

\begin{figure}[!tbp]
  \centering
  \subfigure{\includegraphics[width=0.24\textwidth]{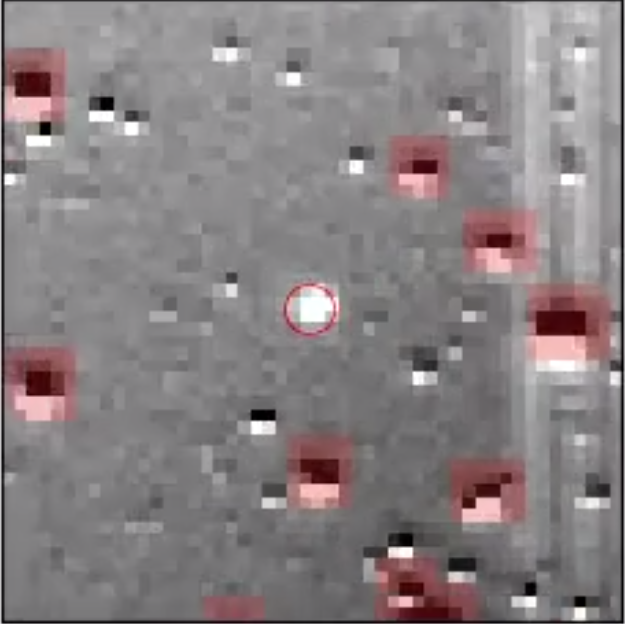}\label{fig:f1}}
  \subfigure{\includegraphics[width=0.24\textwidth]{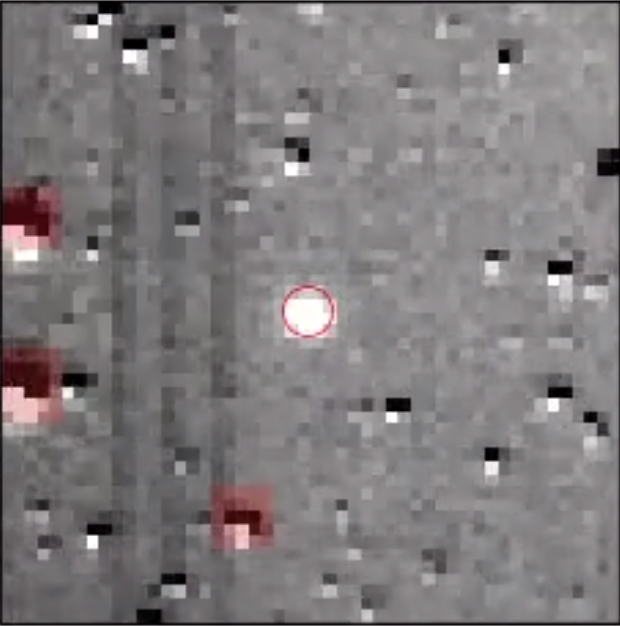}\label{fig:f2}}
  \subfigure{\includegraphics[width=0.24\textwidth]{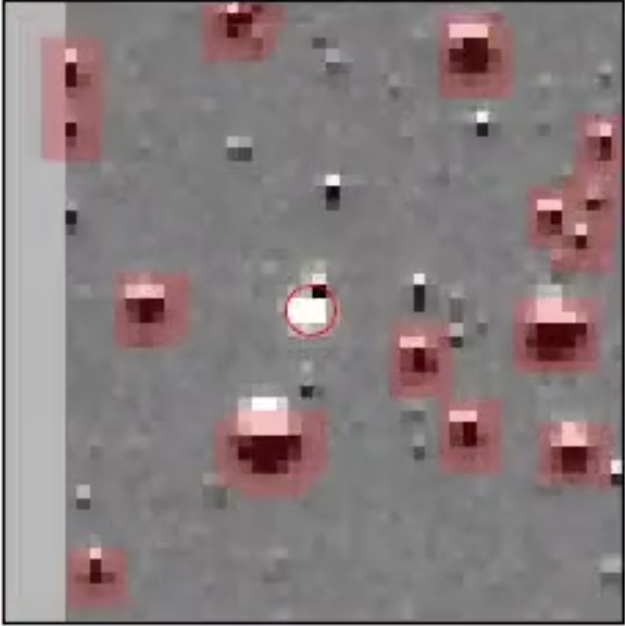}\label{fig:f3}}
  \caption{Three images showing asteroid (1693) Hertzsprung detected by Camera 1 at $t = 0, 12, 24$ days in (a)-(c), respectively. The images have been background subtracted and masked; masked areas are highlighted as red areas. A full animation showing this object and ``real time'' construction of its lightcurve can be found at \url{https://vimeo.com/323253379} .}
  \label{mommertgif}
\end{figure}

The second step  obtains highly accurate ephemerides for each of the candidate asteroids. Ephemerides for each candidate were obtained using {\tt astroquery.jplhorizons} \citep{ginsburg2019} from the JPL Horizons service accounting for the actual location of TESS and the actual epoch of the observations. Rejecting asteroids with positions outside the field of view of the CCD, visual magnitudes less than 20, and positional uncertainties greater than 1\arcsec, this method builds a catalog of asteroids that are most likely to be found in the corresponding CCD. Finally, forced aperture photometry is performed on the accurate positions of asteroids identified by the aforementioned procedure. Figure~\ref{mommertgif} shows sample thumbnails for asteroid (1693) Hertzsprung.

\section{Lightcurves from TESS}

A commonly used method to derive a best-fit period from time-series data is the Lomb-Scargle analysis technique (\citealt{lomb1976}, \citealt{scargle1982}). The main shortcoming of the Lomb-Scargle periodogram is that it does not account for uncertainties in the input data. As such, we use the ``Generalised Lomb-Scargle Periodogram'' (\citealt{zechmeister2009}), which takes these uncertainties into account. This method also fits for the mean of the observed data rather than simply assuming that it is identical to that of the fitted sine curve, an assumption made by the initial form of the LS periodogram. Over a range of rotational frequencies, $\omega$, we obtain the best fit to the observed data through chi-squared minimisation. Assuming that the variation in the asteroid lightcurves is due to rotation of the object we assume the most likely period to be twice the best fit period from our Lomb-Scargle periodogram. This will produce the expected double peaked lightcurve from rotation. To estimate the uncertainty in the period obtained we use Equation~\ref{eqn:period_uncertainty} given in \cite{horne1986} from the work of \cite{kovacs1981}:

\begin{equation}
\delta\omega = \frac{3\pi\sigma_{N}}{2\sqrt{N}TA}
\label{eqn:period_uncertainty} .
\end{equation}

In this equation $A$ is the lightcurve amplitude, $T$ is the total time spread of the dataset, and $\sigma_{N}$ is the variation of the noise in the data. Thus, from our data for each asteroid we derive best-fit periods and amplitudes, both with uncertainties. The uncertainty in light curve amplitude will be due to the photometric uncertainty of the data itself.

\begin{figure}
  \begin{center}
\includegraphics[width=0.45\textwidth]{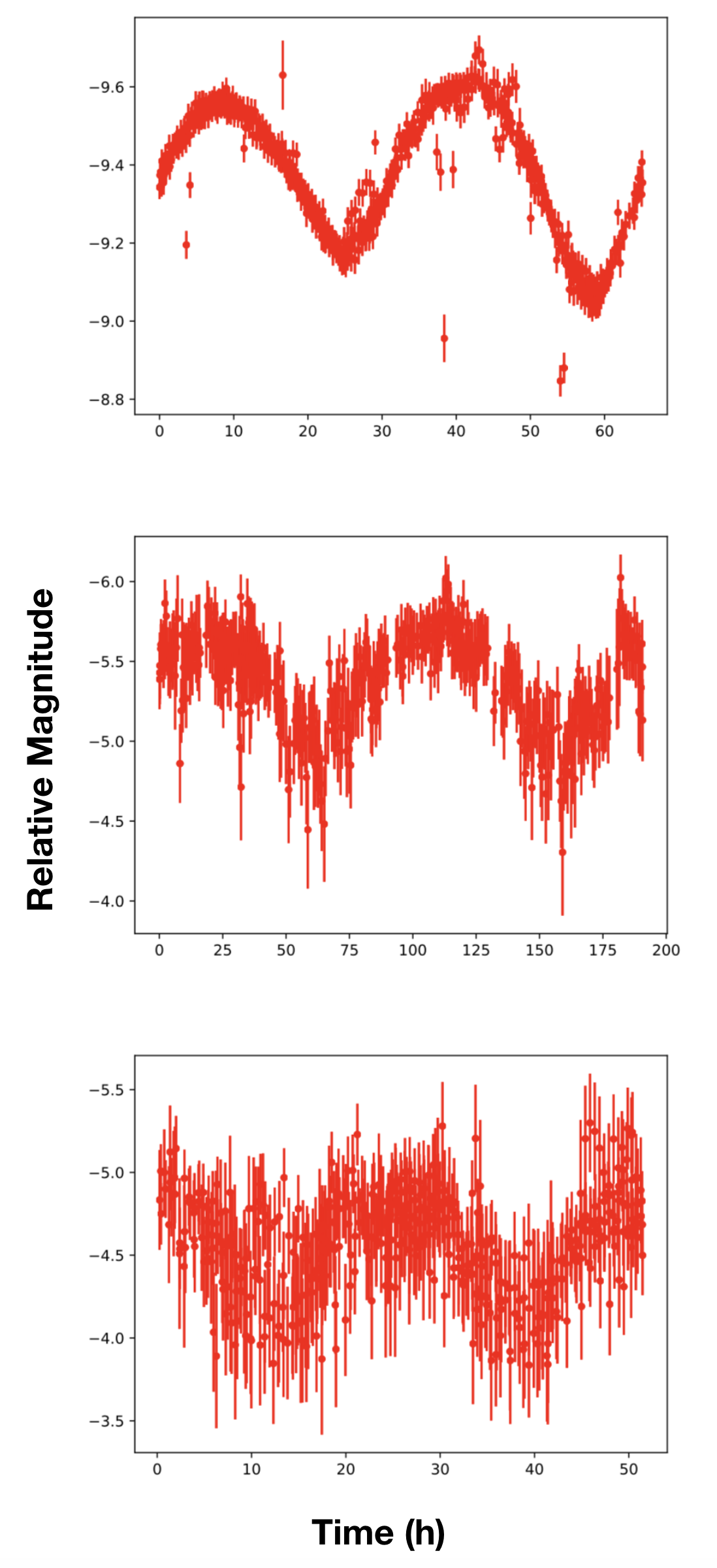} 
\caption{Lightcurves obtained
for asteroids
486, 62561, and 182329 (top to bottom),
which have approximate V magnitudes of
14.4, 18.0, and 18.4, respectively.
The derived rotation periods for these objects are $65.2 \pm 0.1$ h, $191.8 \pm 0.5$ h and $51.50 \pm 0.06$ h, respectively. The lightcurves are phased to their best fit rotation periods.}
\label{tesslc}
  \end{center}
 \end{figure}

From sectors 1+2 we have obtained full lightcurves for 300 objects for which the rotation period has been derived and a limit on the elongation of the bodies set. Comparing objects for which we derive periods with data stored in the LCDB we determine the best cut-off in normalised signal strength for the Lomb-Scargle periodogram here to be $0.4$. Decreasing this value leads to more false positive periods and increasing this cut-off reduces the number of rotation periods derived, including known values for which we derive an identical answer to within uncertainties.
A representative sample of TESS lightcurves is displayed in Figure~\ref{tesslc}.


 \begin{figure}
  \begin{center}
\includegraphics[width=0.5\textwidth]{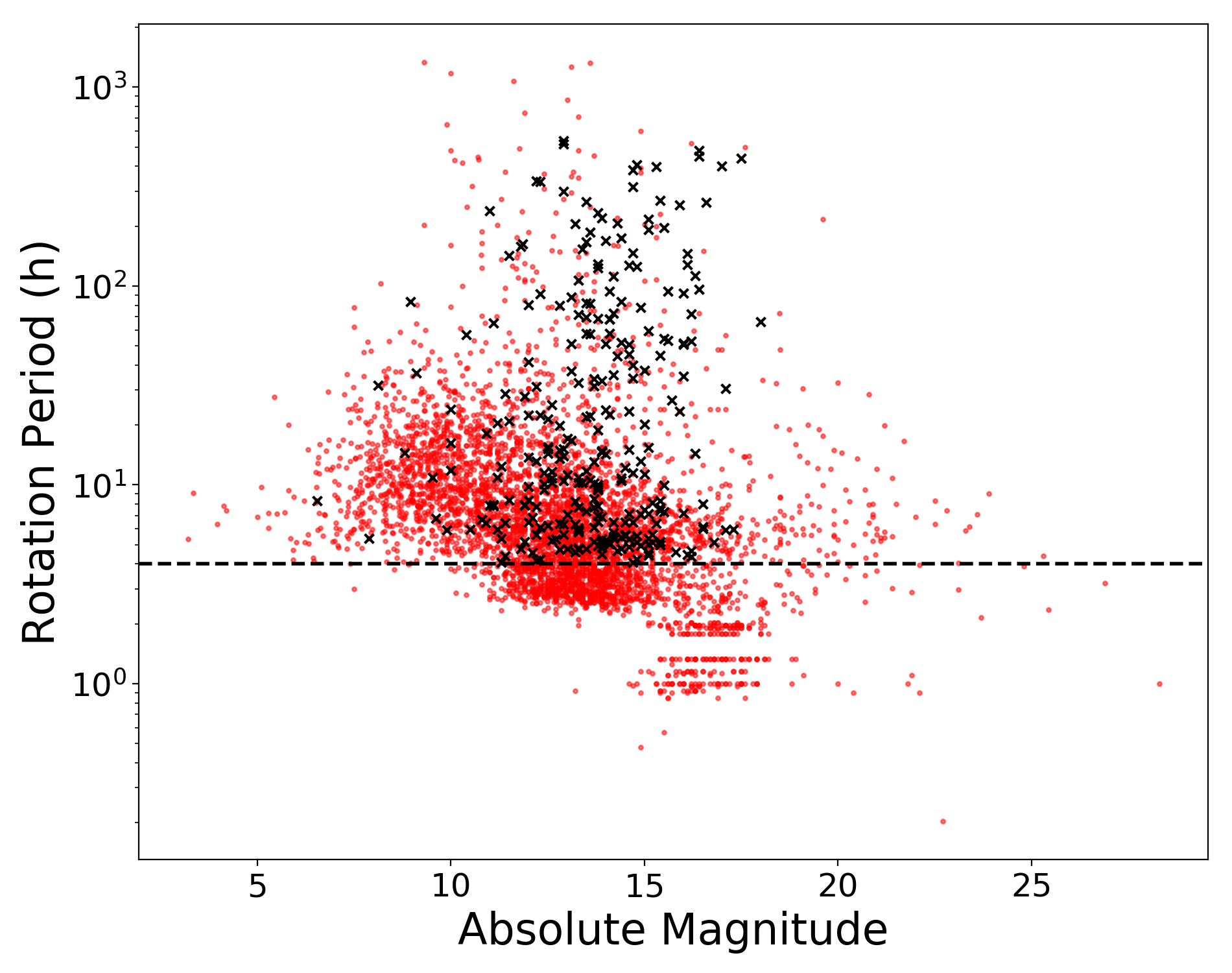} 
\caption{A comparitive plot of derived rotation period against absolute magnitude for data from the LCDB (red data points) and those periods found in this work (black points). The dashed line shows the lower limit on period sensitivity in this data, approximately 4 hours.
}
\label{tesslcdb}
  \end{center}
 \end{figure}

\subsection{Overall results}

From TESS sectors 1 and 2, we find complete period solutions for 300 asteroids. 
The derived periods for these bodies range from $\sim4$ hours to $\sim26$ days with 43 asteroids having rotation periods $P > 100$ h.
The derived rotation periods are overplotted on existing measurements for MBAs from the LCDB in Figure~\ref{tesslcdb}.
We also find partial lightcurves for 7277 asteroids; this result is discussed below.

\begin{figure}
  \begin{center}
\includegraphics[width=0.45\textwidth]{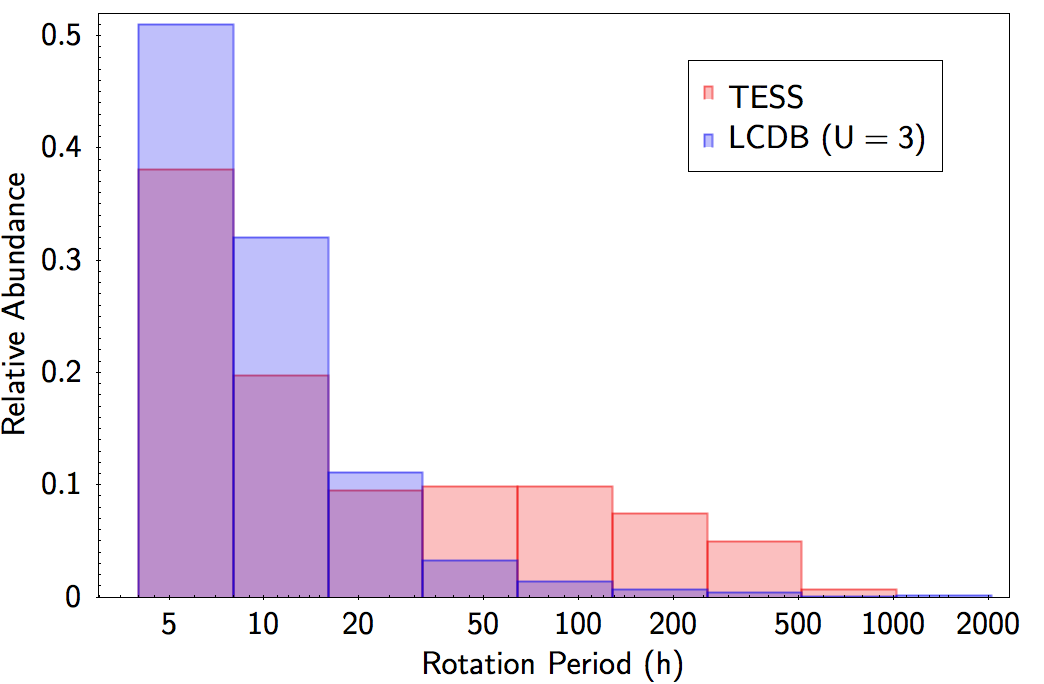}
\caption{A comparison of all of the derived rotation periods from TESS data with existing literature values stored in the Lightcurve Database (e.g., Figure~\ref{lcdb_histogram}). For comparison, we only consider LCDB periods $P > 4$ h, the same limiting period to which we are sensitive in our period determinations. It can be seen in this plot that we have derived rotation periods for proportionally more slow rotators in our sample than the overall known sample to date stored in the LCDB. Our lower limit in period sensitivity can also be seen by the dearth of fast rotators in the sample.}
\label{tesslcdbhist}
  \end{center}
 \end{figure}
 
The distribution of these complete periods is shown in Figure~\ref{tesslcdbhist}.
 We do not have a sufficient sample of objects with $D > 40$ km to reproduce the known Maxwellian frequency distribution for large asteroids.
 However, for the objects with $D < 40$ km in our data set we clearly see an excess of slow rotators relative to a Maxwellian distribution. We do not observe the expected excess of fast rotators relative to a Maxwellian distribution because of the lower limit on period sensitivity of this data (approximately $P=4$ h). We have used the Kolmogorov-Smirnov test to verify that the LCDB and TESS period distributions are truly different and find that this is the case.

\subsection{Comparison with known rotation periods}

\begin{figure}
  \begin{center}
\includegraphics[width=0.45\textwidth]{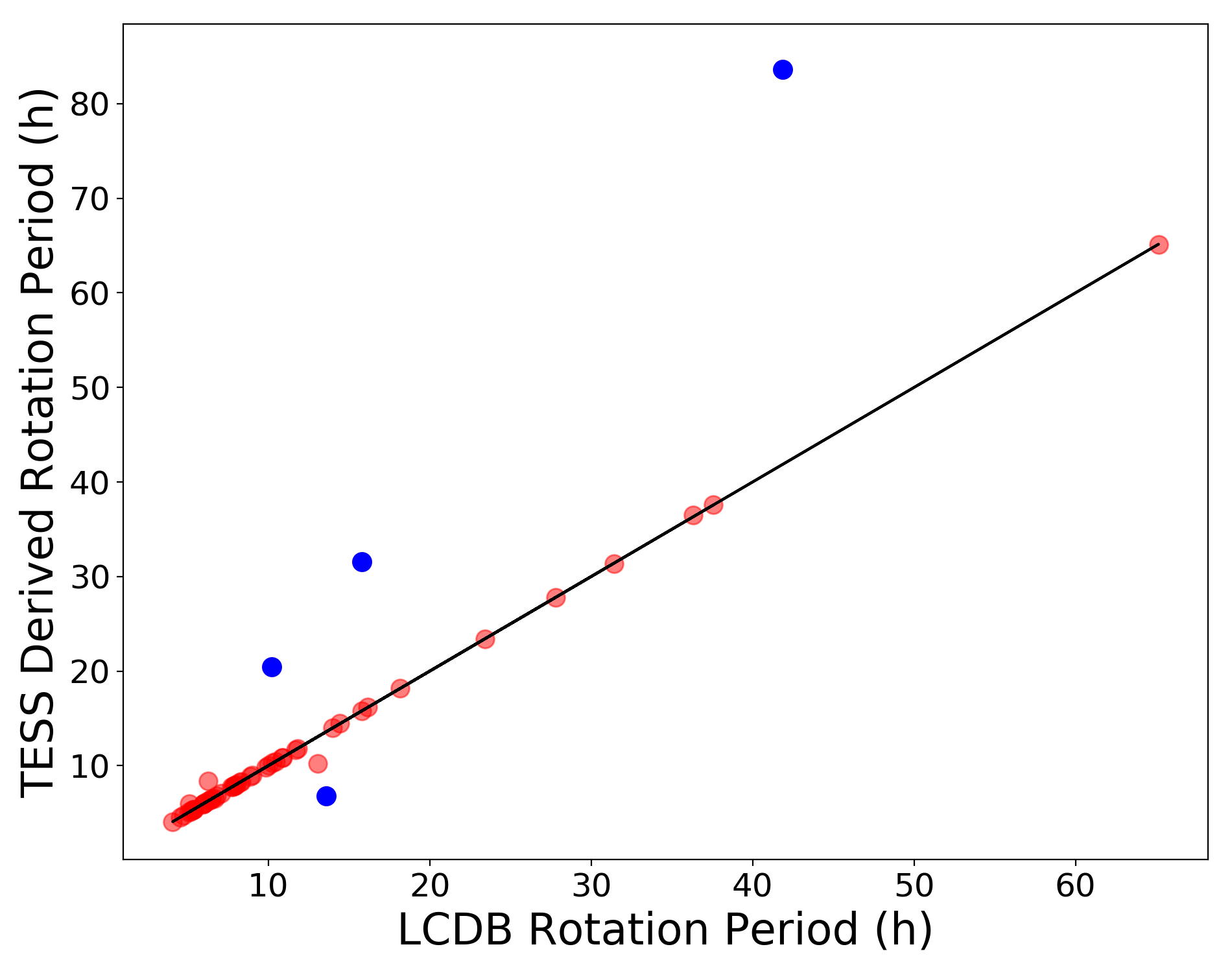} 
\caption{A comparison of the derived rotation periods from TESS data with existing literature values stored in the Lightcurve Database. Red points indicate all objects in our sample with existing rotation periods stored in the LCDB with quality code, U = 3.
The black line represents a 1:1 ratio between the two values and blue points denote objects for which we obtained a best fit for a period different from the literature value by a factor of 2.}
\label{tesslcdbperiods}
  \end{center}
 \end{figure}

Of the 300 rotation periods derived in this initial work, 55 already have rotation periods in the LCDB with quality code $U=3$ suggesting an unambiguous period solution. Of these 55 
our period solutions agree with the previously
determined values, within errors, for 48~asteroids.
Additionally, four of our asteroids have period solutions that are different from the literature value by a factor of~2, suggesting
that one solution or the other is affected by harmonics of the true period.
In only three cases the period determined was different than the LCDB value. Figure~\ref{tesslcdbperiods} shows the determined values for these 55 asteroids plotted against the current values stored in the LCDB.

Most asteroids in this dataset for which rotation periods were already unambiguously known are short period objects, with only 4 
(7\%)
having listed periods $P > 30$ h and none with $P > 100$ h. 

\subsection{Elongated asteroids and intrinsic strength}

For any individual lightcurve we correct the measured amplitude for phase angle effects and use this equation to determine a lower limit on the elongation of the asteroid. The variation in the lightcurve can be used as a means of estimating the axis ratios of the asteroid, as shown in Equation~\ref{eqn:deltam} where $A_{obs}$ represents the observed amplitude of the light curve, $\theta$ is the latitude of the spin pole axis and the three axes of the asteroid are represented as $a$, $b$, and $c$. Assuming the spin axis of the object to be perpendicular to the orbital plane i.e. $\theta=90^{\circ}$, this reduces to Equation~\ref{eqn:deltam2}.

\begin{equation} 
A_{obs}=-2.5log\frac{b}{a}\sqrt{\frac{a^{2}cos^{2}\theta + c^{2}sin^{2}\theta}{b^{2}cos^{2}\theta+c^{2}sin^{2}\theta}}
\label{eqn:deltam}
\end{equation}

\begin{equation} 
A_{obs}=-2.5log\frac{b}{a}
\label{eqn:deltam2}
\end{equation}

To constrain the potential cohesive strengths of these rotating ellipsoids we use a simplified form of the Drucker-Prager model, a stress-strain model commonly used in the study of geological materials (\citealt{alejano2012}). The Drucker-Prager failure criterion is a three-dimensional model estimating the stresses within a geological material at its critical rotation state. The shear stresses on a body in three orthogonal $xyz$ axes are dependent on the density of the body, $\rho$, its rotational frequency, $\omega$, and the major axes of the object $a$, $b$ and $c$ (\citealt{holsapple2007}).

\begin{equation} 
\sigma_{x}=(\rho\omega^{2}-2\pi\rho^{2}GA_{x})\frac{a^{2}}{5}
\label{eqn:sigmax}
\end{equation}
\vspace{-0.3cm}
\begin{equation} 
\sigma_{y}=(\rho\omega^{2}-2\pi\rho^{2}GA_{y})\frac{b^{2}}{5}
\label{eqn:sigmay}
\end{equation}
\vspace{-0.35cm}
\begin{equation} 
\sigma_{z}=(-2\pi\rho^{2}GA_{z})\frac{c^{2}}{5}
\label{eqn:sigmaz}
\end{equation}

The three $A_{i}$ functions are dimensionless parameters dependent on the axis ratios of the body (more specific details on these parameters
are given in \cite{mcneill2018}). The Drucker-Prager failure criterion defines the point at which the object will break up. Using a simple model based on this failure criterion we determine the required cohesive strength as a function of density for each observed object.
We use a Monte Carlo numerical simulation to determine solutions for a range of values using the uncertainties in each of the asteroid parameters to constrain required cohesion of each object.

\begin{figure}
  \begin{center}
\includegraphics[width=0.45\textwidth]{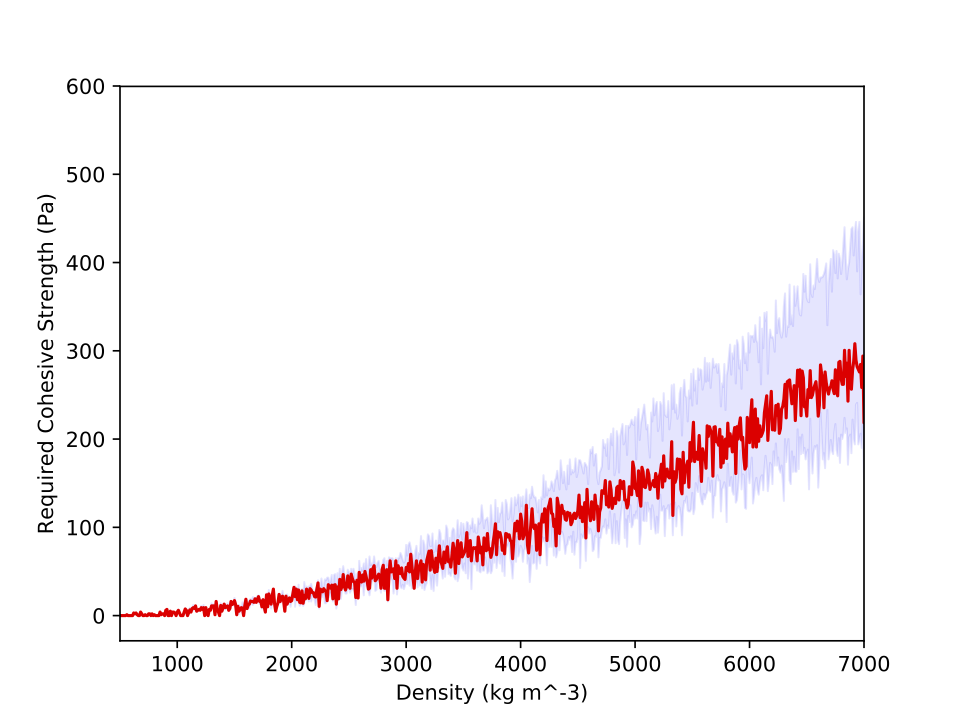}
\caption{The density-dependent strength of asteroid 258045 calculated using the derived rotation period from TESS $P = 112.9 \pm 1.4$ h and a phase-corrected amplitude $A = 1.54 \pm 0.30$ mag.
}
\label{tessstrength}
  \end{center}
 \end{figure}

From TESS data we have determined asteroid 258045 to have a rotation period $112.9 \pm 1.4$ h with a phase-corrected amplitude $1.54 \pm 0.30$ mag. The density-dependent strength profile for this object is given in Figure~\ref{tessstrength}. Even at this elongation we find that for the average densities of S and C-type asteroids, approximately the $2000-3500$ kg m$^{-3}$ range,  that only minimal strength is required to resist rotational fission.

\cite{mcneill2018b} define an "extreme asteroid" as an object rotating with a period shorter than the spin barrier  (P < 2.2 h), or an object with measured light curve amplitude $A > 1.0$ mag;
it is this second criterion that we are sensitive to in this work.
Assuming asteroids to be rubble piles, triaxial ellipsoids elongated enough to produce such an amplitude would be unstable due to rotational fission. Therefore, an asteroid with this rotation state must have some degree of cohesive internal strength. Under nominal assumptions, objects with periods longer than 30 hours with amplitudes larger than 1.6 mag would require strength.
From the initial two sectors of TESS data we have not identified any slow rotating asteroids exhibiting large amplitudes such that non-zero intrinsic strength is required.

\subsection{The evolved rotation state of (2320) Blarney}

(2320) Blarney is a D-type main belt asteroid
with a diameter of some 36~km
\citep{neowise}
and a previously determined rotation period $P=5.097 \pm 0.001$ h (\citealt{usui2011}, \citealt{bennefeld2009}). From TESS photometry we have derived a rotation period $5.99 \pm 0.01$ h for this object. A comparison of the two light curves is given in Figure~\ref{comparelc}.

\begin{figure}[!tbp]
  \centering
  \subfigure{\includegraphics[width=0.42\textwidth]{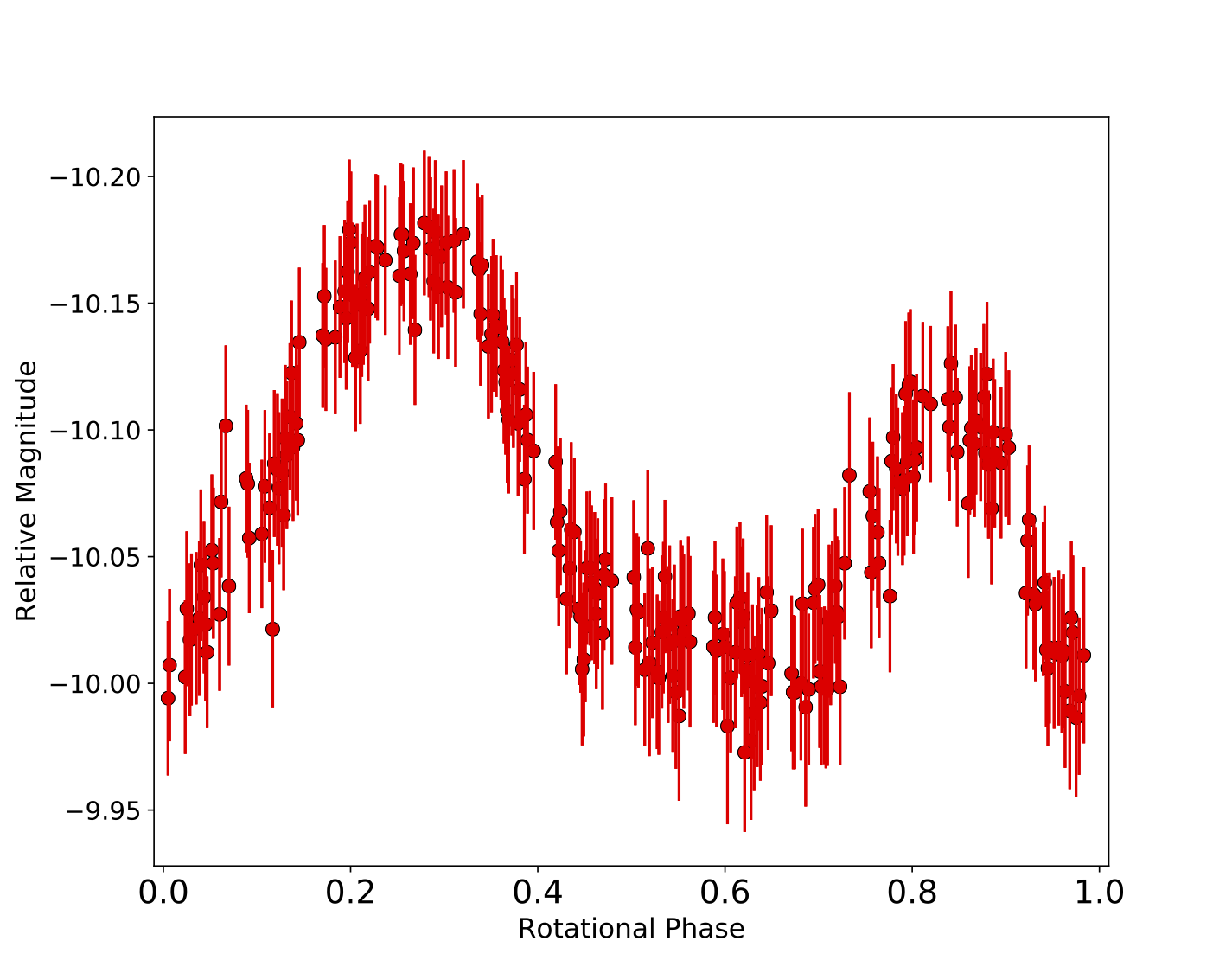}\label{fig:lcf1}}
  \subfigure{\includegraphics[width=0.42\textwidth]{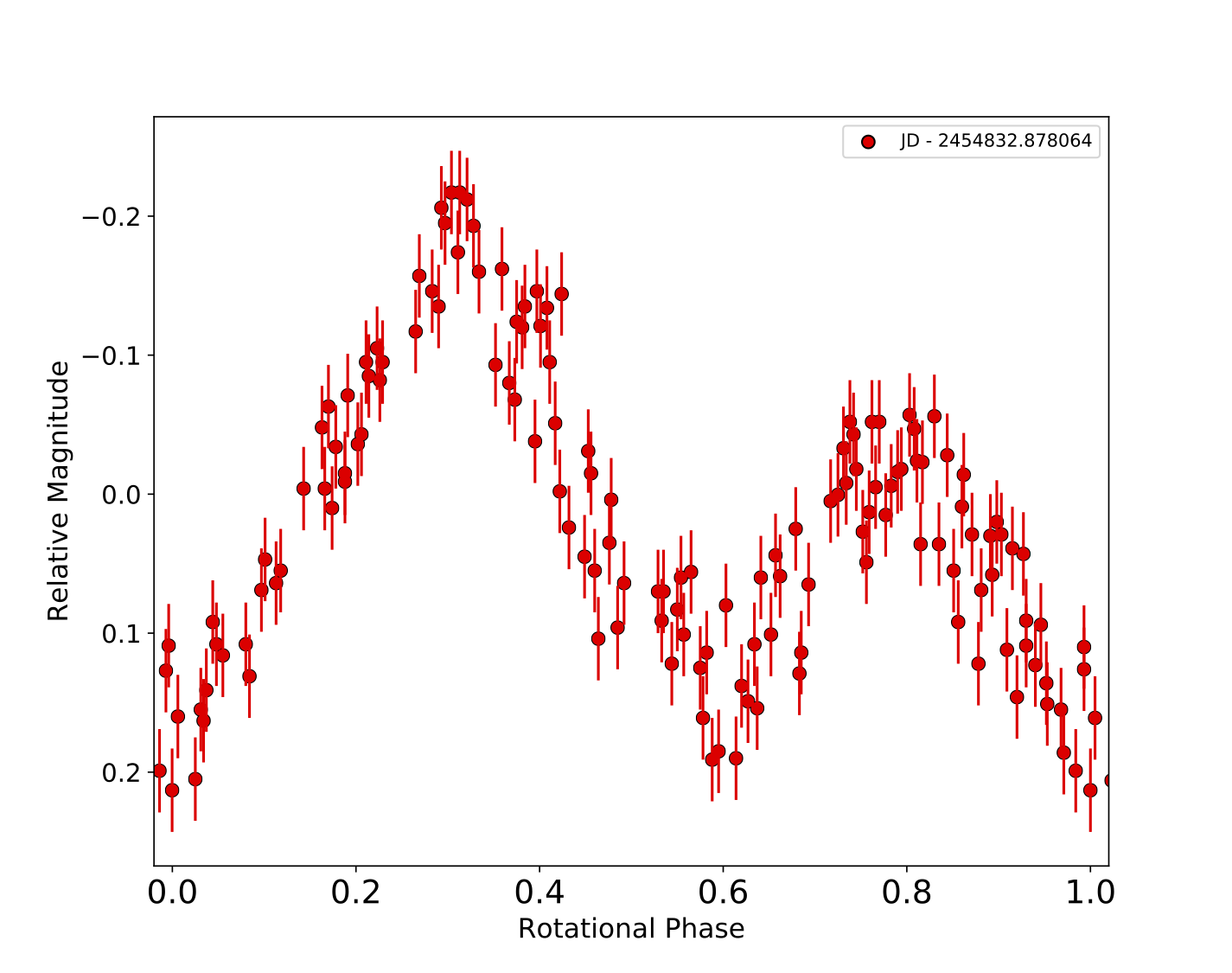}\label{fig:lcf2}}
  \caption{The top panel shows the folded lightcurve obtained from the TESS data with a best fit period $5.99 \pm 0.01$ h. The bottom panel shows a representation of the 2009 lightcurve of (2320) Blarney from \cite{bennefeld2009} with a measured period of $5.097 \pm 0.001$. The given photometric uncertainties are of the order of 0.03 mag for each data point.
}
  \label{comparelc}
\end{figure}

Comparing the two lightcurves the same basic shape can be seen in each case, but the two rotation periods calculated are clearly different. In our periodogram no signal was found corresponding to the period derived in 2009. As a check we folded the TESS data to the \cite{bennefeld2009} rotation period; this produced only scatter. It appears that this object has slowed down by some 20\% over the past decade. We obtained two short (~2.5 hours) pre-dawn CCD photometry
runs on 2320 Blarney using the Lowell 0.7-m telescope
(2019 Aug 27 and 28 UT).  The lightcurve segments are
consistent with a period of $5.98 \pm 0.02$ hours, and show
rms scatter on the fit of 0.012 mag.  Forcing the data to
fit the 5.097 hour period of \cite{bennefeld2009} yields a largely scattered
lightcurve. The main factors influencing evolution of asteroid rotation rates are sub-catastrophic collisions, surface activity, tidal interactions with large bodies, and the Yarkovsky-O'Keefe-Radzievskii-Paddack (YORP) effect. As this is a main belt asteroid, tidal forces due to planetary interactions can be considered negligible. We rule out the YORP effect as the mechanism behind this rotation period change due to its diameter. In the case of a smaller object experiencing YORP, this deceleration would be expected to occur consistently,
and we find that the deceleration of (2320) Blarney corresponds to a value of $\frac{d\omega}{dt} = 1.25 \times 10^{-3}$ rad d$^{-2}$. This is much larger by several orders of magnitude than any previously measured YORP acceleration (e.g., the first direct detection of YORP on asteroid (54509) YORP was found by \cite{lowry2007} to be $3.5 \times 10^{-6}$ rad d$^{-2}$).\\
\\
\textbf{If} the period reported by \cite{bennefeld2009} is correct then this leaves sub-catastrophic  collisions and/or surface  activity as likely mechanisms to change Blarney's rotation rate, and we suggest that a concerted effort to look for activity and/or rotationally varying compositional indicators (colors, spectra) might help inform our understanding of the evolution of this body.

\begin{figure}[!tbp]
  \centering
  \subfigure{\includegraphics[width=0.42\textwidth]{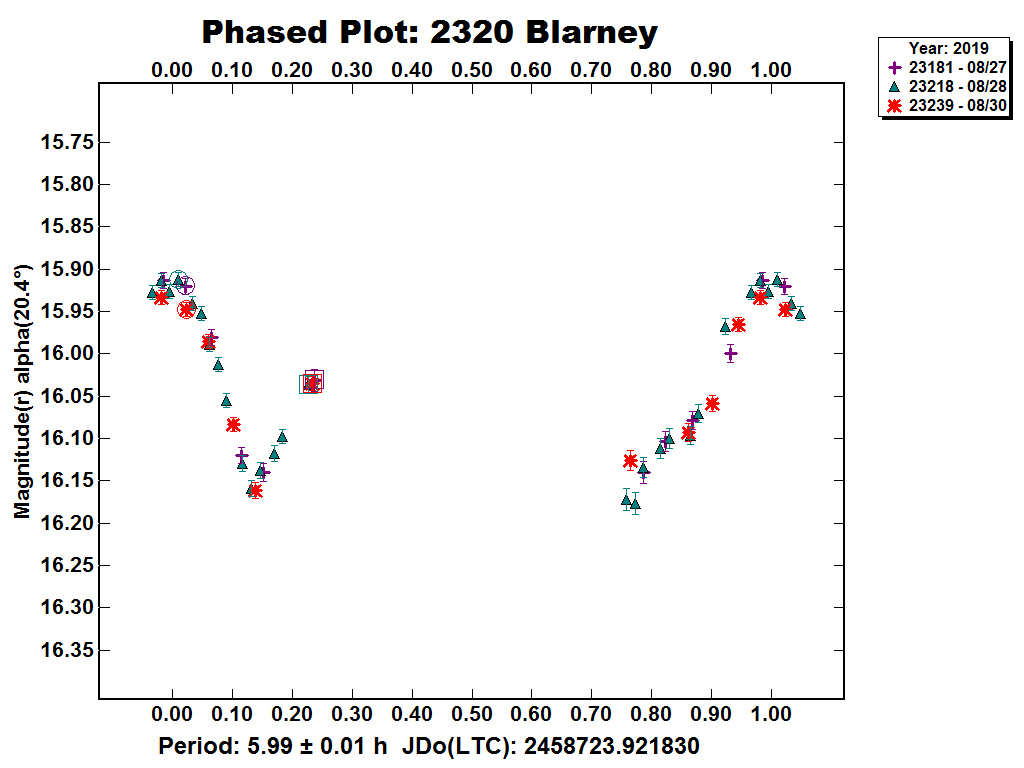}\label{fig:lcLOW1}}
  \subfigure{\includegraphics[width=0.42\textwidth]{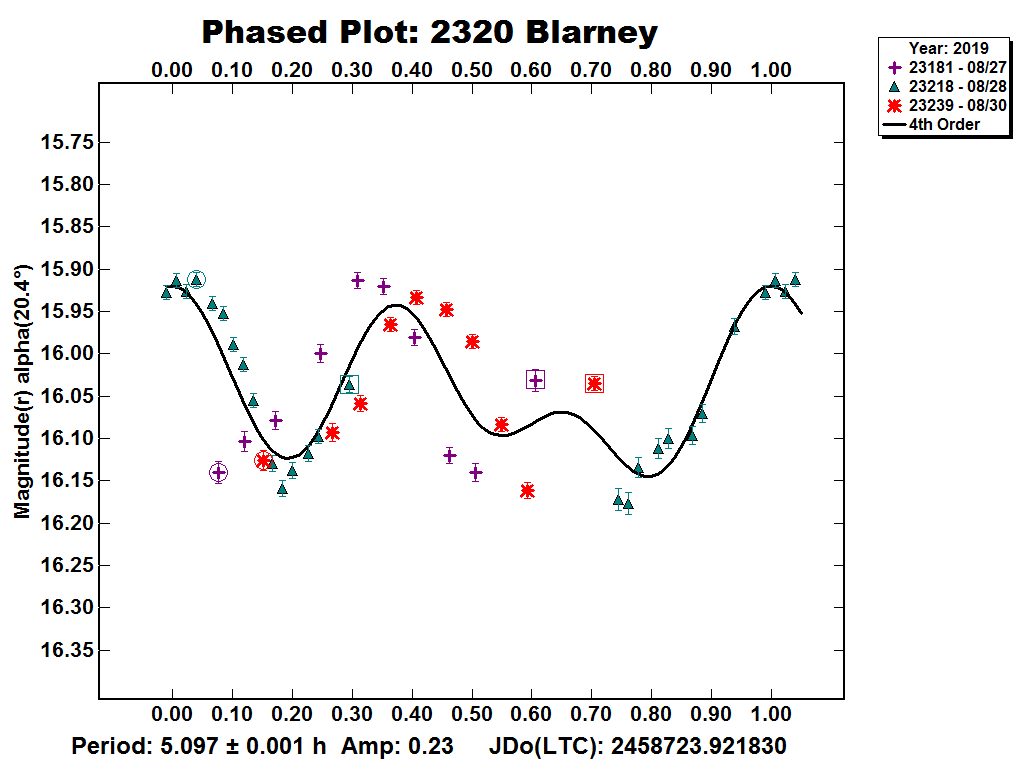}\label{fig:lcLOW2}}
  \caption{The top panel shows the folded lightcurve obtained from the 0.7m Lowell Telescope data with a best fit period $5.99 \pm 0.01$ h. The bottom panel shows the same data folded to the 2009 measured period of $5.097 \pm 0.001$ h.
}
  \label{comparelc}
\end{figure}

\section{Future Work}

We will continue this work as TESS public data releases continue
over the lifetime of the mission, and, if applicable, beyond the nominal mission lifetime should operations continue. The work in this paper is derived from the first two pointings of the TESS mission. Over the nominal lifetime of the mission TESS will observe 26 pointings. Extrapolating the numbers obtained in this pilot study we anticipate obtaining full lightcurves for approximately 6000 main belt asteroids with partial or fragmentary lightcurves for a further 79,000 objects. We expect that $\sim 10\%$ of these objects will have rotation periods $P > 100$ h, which represents an increase of a factor of 10 in unambiguous period solutions for objects in this period range assuming an LCDB quality code of $U \geq 3-$ and a factor of 2 assuming $U \geq 2-$.

Sparse photometry and partial lightcurves can be used to determine an estimate for the shape distribution of a population. Previous work on this has been carried out in \cite{mcneill2016} and \cite{cibulkova2018}. Both used sparse photometry from the Pan-STARRS 1 Survey to obtain a shape distribution for main belt asteroids (MBAs). Both studies were in good agreement that, if the objects are assumed to be a population of prolate spheroids ($a > b \geq c$), the average axis ratio for a small ($1 < D < 10$ km) main belt object is approximately $\frac{b}{a}=0.8$. Neither of these studies accounted for the presence of slow rotators within the data as a reliable period distribution encompassing this range did not exist. With future data releases it will be possible to constrain this shape distribution for main-belt asteroids while also more accurately accounting for the presence of slow rotators in the sample. We expect to produce an initial shape distribution from this data after the public releases of another several sectors.

\section{Conclusions}

We have carried out a pilot study using public TESS data from the mission's first two sectors.
From this 54~day coverage we have derived rotation periods for 300~main-belt objects and have constructed partial or fragmentary lightcurves for a further 7277 asteroids. For the sample of 55 short-period objects which have previously derived periods, we find our rotation periods to be in excellent agreement with the earlier results. 
We are sensitive to rotation periods of
up to $\sim$10~days,
and in this preliminary project we find 43 asteroids with periods $P > 100$ h.  This is significant because traditional classical observing is generally biased against long periods, which means that, until now, our understanding of the distribution of shapes, rotation periods, and strengths for main belt asteroids is incomplete. Of these slow rotators we find none have a combination of rotation rate and elongation requiring any internal strength to resist rotational reshaping.
Over the nominal two-year lifetime of the mission, we expect to obtain detections for 85,000 unique asteroids and expect to construct full light curves for around 6000 of these. Over the mission lifetime we expect an increase of a factor of 10 in unambiguous period solutions for objects in this period range assuming an LCDB quality code of $U \geq 3-$ and a factor of 2 assuming $U \geq 2-$.

In addition, we have identified an object, (2320) Blarney, that has \textbf{potentially} undergone rotational deceleration over the decade since its rotation period was initially determined.  Sub-catastrophic collisions and/or surface activity may have produced this deceleration, and an observational campaign to search for evidence of these effects might help
inform our understanding of this body.

\acknowledgments
\textbf{Acknowledgments}

We thank the anonymous referee for their comments on this manuscript which led to an increase in its overall quality. We thank Colin Chandler for his assistance with large scale JPLHorizons queries. AM was supported in part by the Arizona Board of Regents' Regents Innovation Fund.

\facilities{TESS} 
\software{Numpy (\citealt{numpy}), Astropy (\citealt{astropy}), astroquery \citep{ginsburg2019}, sbpy \citep{mommert2019}, photutils \citep{photutils}}



\begin{deluxetable}{ccccccc}
\tablecaption{Summary table of rotational information derived from pilot-study TESS observations. Normalised peak signal is the peak periodogram value when the full data set is analysed i.e. no additional cropping or sigma-clipping. (Uncertainties included here are provisional and should be considered to be conservative, full uncertainties and further information will be included in a subsequent database release.)
}
\tablehead{\colhead{Object} & \colhead{H} & \colhead{Coverage of Obs. (h)} &\colhead{Rot. Period} & \colhead{Rot. Uncertainty}& \colhead{Normalised Peak Signal} & \colhead{Observed Amplitude}}
\startdata
\input{lcdata.txt}
\enddata
\label{table:obs}
NOTE - Table 1 is published in its entirety in the machine-readable format. A portion is shown here for guidance regarding its form and content
\end{deluxetable}


\end{document}

%% file: lcdata.txt
70&8.11&668&31.551&0.012&0.41&$0.09\pm0.04$\\
196&6.54&668&8.339&0.001&0.53&$0.09\pm0.02$\\
327&9.9&665&5.937&0.001&0.70&$0.33\pm0.05$\\
416&7.89&209&5.370&0.002&0.81&$0.27\pm0.03$\\
436&10&52&16.178&0.159&0.76&$0.19\pm0.05$\\
481&8.8&657&14.456&0.002&0.60&$0.13\pm0.09$\\
486&11.1&495&65.178&0.074&0.88&$0.49\pm0.05$\\
498&8.95&603&83.549&0.098&0.41&$0.08\pm0.03$\\
582&9.11&657&36.489&0.013&0.41&$0.10\pm0.04$\\
598&9.53&332&10.895&0.011&0.82&$0.40\pm0.11$\\